\documentclass[12pt]{iopart}
\pdfoutput=1
\usepackage{iopams}
\expandafter\let\csname equation*\endcsname\relax

\expandafter\let\csname endequation*\endcsname\relax

\usepackage{citesort}

\usepackage{array}
\usepackage{multirow}
\usepackage{multicol}
\usepackage{ulem}
\usepackage{amsmath}
\usepackage[a4paper, total={6in, 10in}]{geometry}
\usepackage[numbers]{natbib}
\usepackage{graphicx}
\usepackage[utf8]{inputenc}
\usepackage[T1]{fontenc}
\usepackage{physics}
\usepackage[breaklinks=true,colorlinks=true,linkcolor=blue,urlcolor=blue,citecolor=blue]{hyperref}
\hypersetup{
	colorlinks   = true,
	citecolor    = cyan,
	urlcolor     = red,
}
\usepackage[usenames, dvipsnames]{color}
\usepackage[mathscr]{euscript}
\usepackage[T1]{fontenc}
\usepackage[font=small,labelfont=bf,tableposition=top]{caption}

\DeclareCaptionLabelFormat{andtable}{#1~#2  \&  \tablename~\thetable}

\begin{document}


	\title{A protocol to create a multi-particle entangled state for quantum-enhanced sensing.}
	
	\author{Rahul Sawant~\footnote{Present Address: M Squared Lasers Ltd, 1 Kelvin Campus, West of Scotland Science Park, Maryhill Road, Glasgow, G20 0SP, UK}}
	\address{School of Physics and Astronomy, University of Birmingham, Edgbaston, Birmingham B15 2TT, United Kingdom.
	}%
	\eads{\mailto{r.v.sawant@bham.ac.uk, rahul.sawant@m2lasers.com}}

	\date{\today}

	\begin{abstract}
		We propose a protocol for generating multi-particle entangled states using coherent manipulation of atoms trapped in an optical cavity. 
		We show how entanglement can be adiabatically produced with two control beams and by exploiting cavity-mediated interactions between the atoms. Our methods will allow for optimal generation of entanglement for the measurement protocol we propose. We discuss an experimental implementation and compare the performance of the states produced with those of classical states and ideal maximally-entangled Dicke states. We find that our states always feature metrological gain and even outperform ideal Dicke states in the measurement of magnetic field gradients. Due to the easy scalability, our entanglement protocol is a promising tool for quantum state engineering.   
	\end{abstract}
	\maketitle
	
	\section{\label{sec_intro}Introduction}
	
	Devices exploiting the laws of quantum mechanics promise to deliver the next step-change in technology.
	In particular, quantum sensing exploiting multi-particle quantum entanglement could have a substantial impact in our daily lives, as it features better scaling of precision compared to classical methods~\cite{Pezze_Rev_2018, Monz_2011,berrada_integrated_2013,hosten_measurement_2016}. 
	Several classes of multi-particle entangled states have been proposed and realized, among them squeezed states \cite{Meyer_experimental_2001, riedel_atom_2010, Wu_retrieval_2020}, Dicke states \cite{barontini_deterministic_2015, Lucke_detecting_2014}, and GHZ or ``Schr\"odinger's cat'' states \cite{leibfried_toward_2004,Monz_2011, omran_generation_2019}. These have been realized in a variety of systems, and among them, interfaces of atoms with optical cavities has delivered so far some of the best performance in terms of metrological gain \cite{haas_entangled_2014,barontini_deterministic_2015,hosten_measurement_2016,Gietka_Supersolid_2019,Torre_Dissipative_2013,Victoria_Probing_2019,Hamilton_atom_2015,Sabulsky_multi_2022}.	
	
	The main issue with entangled states is that they are extremely fragile, and this is particularly marked for symmetric entangled states that require all the atoms in the ensemble to be indistinguishable. This is because it is very difficult to maintain symmetry and indistinguishably in presence of decoherence and particle loss. Asymmetric entangled states are instead more resilient, but they are usually considered less-performing and also difficult to realize in a controlled fashion. In this work, we discuss a novel architecture to optimally generate metrologically useful multi-particle entangled states using atoms trapped in a cavity. In our system, single atoms are trapped in optical tweezers and their interaction is mediated by the field in the cavity. We show that with the addition of two ``transverse" control beams, it is possible to adiabatically generate asymmetric entangled states that are optimised for a certain measurement protocol. We concentrate in particular on the measurement of an external magnetic field and a magnetic field gradient. We compare the performance of the states produced with classical states and ideal maximally-entangled Dicke states. We find that in the regime of parameters explored, our states always outperform classical states and even outperform the ideal Dicke states when measuring a magnetic field gradient. We also show that our protocol can be easily scaled up.
	\begin{figure}
		\centering
		\includegraphics[width=0.55\textwidth]{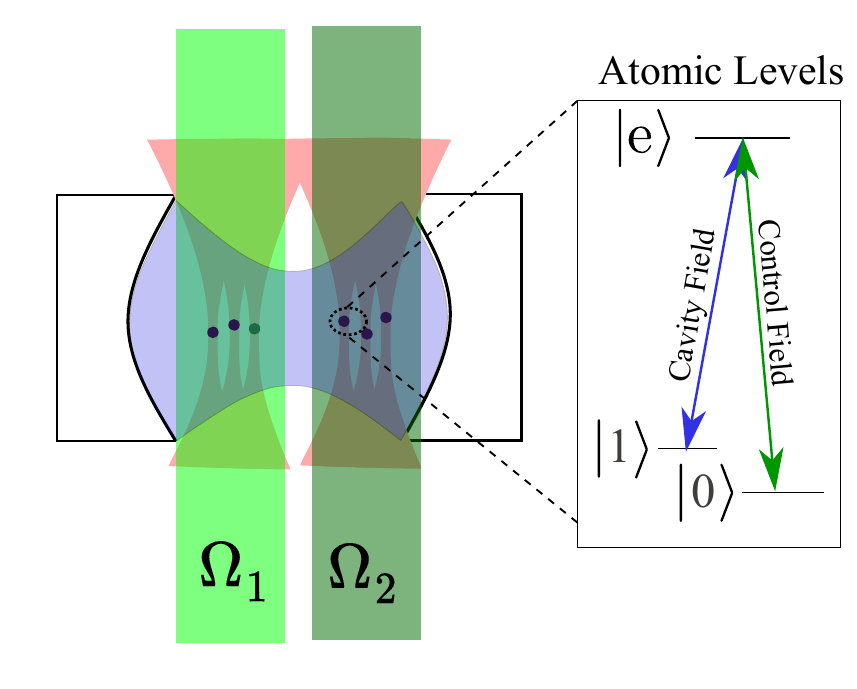}
		\caption{Pictorial representation (not to scale) of the system: individual atoms (black spheres) are trapped in optical tweezers (pink beam). These atoms are interacting with a single mode of a Fabry-Pérot cavity (blue field). The optical tweezers can be generated using a microscope objective. The right of the image shows a representation of a simple 3-level atomic system and the light fields coupling to the transitions. The two control fields (green fields) $\Omega_1$ and $\Omega_2$ couple to two separate halfs of the atoms during the adiabatic creation of the entangled states.}
		\label{fig:cartoon}
	\end{figure}
	
	The article is organized as follows: in Section~\ref{sec:thesystem} we describe the experimental system and we discuss a realistic implementation of it, in Section~\ref{sec_entangled} we introduce the adiabatic protocol to generate optimally-entangled states for the measurement of an external magnetic field and a magnetic field gradient. 
	In section~\ref{sec_sensing}, we quantify the performance of our entangled states and compare their precision with classical states and entangled Dicke states. Section~\ref{conclusion} is finally devoted to the conclusions. 
	
	
	\section{The System}
	\label{sec:thesystem}
	
	Our system is depicted in Fig.~\ref{fig:cartoon} and consists of $N$ individual three-level atoms trapped with optical tweezers inside the mode of an optical cavity.  We divide the atoms into two halves, and each half is individually addressed by dedicated light beams with Rabi frequencies $\Omega_1(t)$ and $\Omega_2(t)e^{i \phi}$ respectively. The cavity and each of the control beams are tuned to be resonant with the atomic transitions as shown in Fig.~\ref{fig:cartoon}. Such a configuration realizes a $\Lambda$-type system for each atom, in which the two ground states act as qubit states. The cavity is used to mediate the interaction between the individually trapped atoms. Trapping atoms in tweezers allows us to suppress detrimental atom-atom collisions.
	
	Our protocol for the generation of multiparticle entanglement consists in finding the optimal evolution for the control light fields $\Omega_1(t)$ and $\Omega_2(t)e^{i \phi(t)}$ such that the Classical Fisher information ($\mathcal{F}$) for sensing a certain parameter $\theta$ is maximum. Detailed derivation of $\mathcal{F}$ for our system is given in~\ref{app:fish_info}. As the discussion here is specific to a feasible measurement protocol, we use Classical Fisher information instead of Quantum Fisher information. Generally, it is hard to find measurement protocols which saturates the bounds given by Quantum Fisher information, hence we avoid it here. For detailed discussion on Quantum and Classical Fisher information see reference~\cite{Pezze_Rev_2018}.
	The Classical Fisher information is related to the measurement uncertainty of an an unknown quantity $\theta$ via the following identity,
	\begin{equation}
	\Delta \theta \ge \sqrt{\frac{1}{M \mathcal{F}}},
	\end{equation}
	where $M$ are the number of measurements. 
	For large $M$, the above identity is known to be saturated by Maximum-Likelihood method~\cite{Braunstein_1992}.

	The evolution of the system density matrix is governed by the following master equation,
	\begin{equation} 
	\label{eqn:master_eqn}
	\begin{split}
	&\frac{d\hat{\rho}}{dt} = - i [\hat{H}_f, \hat{\rho}(t)] - \sum_{A} \sum_{j = 0,1}  \frac{\Gamma_j}{2} \left(\hat{\sigma}^\dagger_{A,j} \hat{\sigma}_{A,j} \hat{\rho} - 2\hat{\sigma}_{A,j}\hat{\rho} \hat{\sigma}^\dagger_{A,j} + \hat{\rho}\hat{\sigma}^\dagger_{A,j} \hat{\sigma}_{A,j} \right) \\
	& -  \frac{\kappa}{2} \left(\hat{a}^{\dagger} \hat{a} \hat{\rho} - 2\hat{a}\hat{\rho} \hat{a}^{\dagger}  + \hat{\rho} \hat{a}^{\dagger} \hat{a}  \right),
	\end{split}
	\end{equation}
	where the Hamiltonian $\hat{H}_f$ takes into account the coupling to the cavity and a classical light field. $A$ denotes a particular atom, $\hat{\sigma}_{A,j} = \ket{j_{A}}\bra{e_{A}}$, $\Gamma_0$ and $\Gamma_1$ are decay rates from level $\ket{e}$ to levels $\ket{0}$ and $\ket{1}$ respectively, $\hat{a}$ is the annihilation operator for the cavity field, and $\kappa$ is the cavity field decay rate. The explicit form of $\hat{H}_f$ is given in the~\ref{app:calculations}.
	
	\begin{figure}
		\centering
		\includegraphics[width=0.9\textwidth]{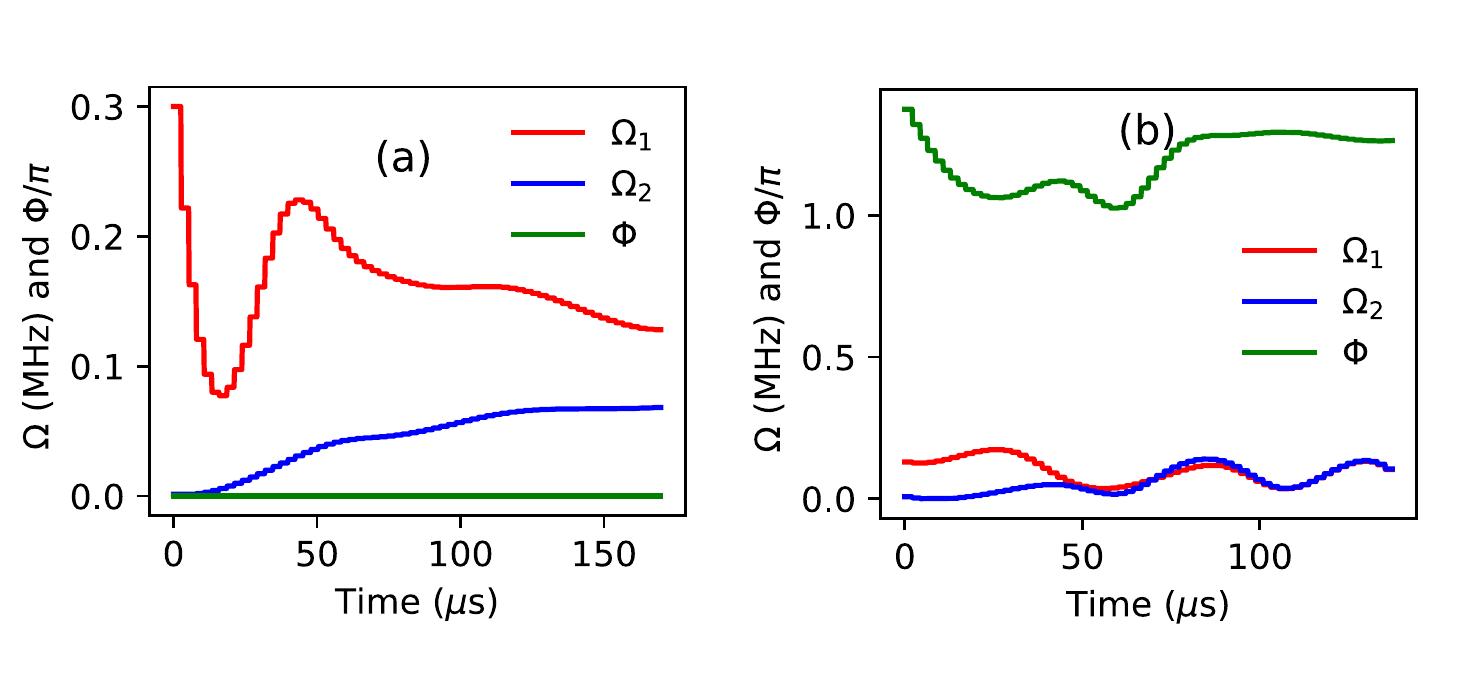}
		\caption{Evolution of the light fields for (a) two atom system for case-I (see section~\ref{sec_sensing}), (b) two atom system for case-II.
		}
		\label{fig:probability_evolution}
	\end{figure}
	
	To illustrate a realistic implementation, we concentrate on Rubidium 87 atoms, which have been the workhorse of cold atoms experiments due to their ease of cooling to ultracold temperatures.  
	As qubit states we choose the $|F,m_F\rangle = |1, 1\rangle$ and $|2, 0\rangle$ hyperfine levels of the $5^{2}\text{S}_{1/2}$ ground state, such that $\ket{0} = |1, 1\rangle$ and $\ket{1} = |2, 0\rangle$. To implement our protocol, it is preferable to minimize off-resonant excitations by separating the transition between the two qubit states from those with other Zeeman states using an external magnetic field. At the same time, it is important to avoid excessive noise coming from magnetic field fluctuations. We found that at a magnetic field of $B_0 = 654.39$ Gauss (see \ref{app:zeeman}), the energy difference between the two qubit states has a low sensitivity to magnetic field noise. Indeed the change in transition frequency between these two levels is $\Delta f \sim  10^{-3}$ Hz for a magnetic field fluctuation of $1$~mG. At the same time, this magnetic field separates the Zeeman states by $\sim 100$~MHz and is easy to produce in an experiment. It is worth noticing that states whose energy difference has low sensitivity for magnetic field noise can also be found for lower magnetic fields~\cite{Szmuk_2015} \footnote{The stable and high magnetic field is only required during the generation of the entangled states. Once these are created, the magnetic field can be adiabatically switched off and a measurement of some other unknown magnetic field can be performed.}. In addition, the energy difference between the levels $|0\rangle$ and $|1\rangle$ does not depend on the intensity of the trapping light as the AC-stark shift for the levels  of $5^{2}\text{S}_{1/2}$ state is independent of $F$ and $m_F$ if linearly polarized light is used for trapping~\cite{Arora_state_2012,le_kien_dynamical_2013}. This will make a quantum superposition of these states insensitive to fluctuations in the intensity of the tweezer trap. The above properties make the chosen levels ideal for the storage of quantum information. It should be possible to extend the coherence time of these states for $> 1$ second if the magnetic field fluctuations are $<100$~mG, which is possible with current technology~\cite{merkel_magnetic_2018,farolfi_design_2019}.  
	
	Concerning the cavity, we choose a common configuration (see e.g. \cite{Welte_photonmediated_2018}) with length 1 mm and radius of curvature of the cavity mirrors 10 mm. This results in $g_0 \sim 4$ MHz, where $\hbar g_0  = \mu_{1,e}\sqrt{\hbar \omega_c/(2  \epsilon_0)}$ is the interaction strength between the cavity field and a single atom~\cite{steck_quantum_2011}, $\omega_c$ is the cavity resonance frequency, and $\mu_{1,e}$ is the transition dipole moment between the $\ket{1}$ and $\ket{e}$ levels \footnote{Here we consider the excited state $\ket{e}$ to be  the eigenstate closest in composition to $(F,m_F) = (1, 0)$ level of the $5^{2}\text{P}_{1/2}$ electronic state (This state is part of the D1 spectrum in $^{87}$Rb~\cite{Steck_Rb_87}). After diagonalizing the hyperfine and Zeeman Hamiltonian at a magnetic field of $B_0 = 654.39$ Gauss, the eigenstate closest to the state $(1, 0)$ is not pure and has a 10$\%$ mixture from other states. We take this into consideration when calculating $\mu_{1,e}$, which lets us accurately calculate $g_0$.}. Considering a reflectivity of $99.9999~\%$~\cite{Cole_16,Maldaner_20} for the cavity mirrors, we obtain a cavity loss rate of $\kappa/\hbar \sim 0.1$~MHz. Our cavity parameters should allow for up to 1000 spatially separated atoms with nearly-identical atom-cavity coupling  ($\ge 0.9 g_0$ for all the atoms), that can be accurately positioned using the optical tweezers~\cite{Endres1024,barredo_synthetic_2018}. The position stability of the atoms can be achieved by cooling the atoms to their vibrational ground state in the tweezer trap. This will localize the atoms to much smaller distance ($\sim$~30 nm) than the wavelength of the light. This can be achieved using techniques demonstrated in reference~\cite{Kaufman_cooling_2012}. 
	\section{Entangled state creation}
	\label{sec_entangled}
	The goal of our protocol is to find the optimal pulse sequence that maximizes $\mathcal{F}$ for two cases. Where case-I is all  atoms sensing a common field, and case-II is the two halves sensing the difference between two fields. At the same time, to minimize decoherence we must avoid populating the state $\ket{e}$. Therefore we consider only quasi-adiabatic pulse sequences~\cite{wu_fast_2017}. The maximization is done using generalized simulated annealing~\cite{TSALLIS_1996}. The optimization variables include one value of the total time $T$ and 10 values each of $\Omega_1(t)$, $\Omega_2(t)$, and $\phi(t)$. The 10 values are equally spaced between $T$. The values between these points are then found using a smooth cubic interpolating function. During the optimization, the points are bounded in range (0,1) MHz for the amplitudes and (0, 2$\pi$) for the phase difference. Unless stated otherwise, we use the parameters $2\Gamma_0 = 2\Gamma_1 = 5.75$~MHz, $g_0 = 4$~MHz, and $\kappa = 0.1$~MHz. To simulate conditions close to the experimental one, we consider that (i) the change in $\Omega_1(t)$ and $\Omega_2(t)$ is step-wise, with each step corresponding to $\sim 1~\mu$s. This is achieved by digitizing the interpolation function found during the optimization. (ii) The control field resolution is 8-bit such that $\Omega$ can take values only in steps of $\Omega_\text{max}/2^8$. (iii) As detuning from excited state, $\Delta$ and the two-photon detuning, $\delta$ (see \ref{app:calculations} for definitions) are noisy variables, they are randomly sampled during each time step from a Gaussian distribution with a mean of zero. The standard deviation for the above Gaussians for $\Delta$ and $\delta$ are 1~MHz and 10~kHz, respectively. For the cavity, we consider only excitations up to $n = 1$. We observe that including higher photon number states does not change $\mathcal{F}$ because during the evolution of the system the cavity is only virtually excited.    
	
	To simulate Eqn.~\ref{eqn:master_eqn} we use the python numerical package Qutip~\cite{qutip2}. As an example, in  Fig.~\ref{fig:probability_evolution} we show the evolution of $\Omega_1(t)$, $\Omega_2(t)$ and $\Phi(t)$ in a two-atom system for case-I and case-II.
	In general, we find that the states produced at the end of the evolution have the form
	\begin{equation}
	\label{eqn:state}
	\hat{\rho}_\text{DL} \approx p\ket{DL_N^{(m)}}\bra{DL_N^{(m)}}+(1-p)\hat{\rho}_\text{loss},
	\end{equation}
	where
	\begin{equation}
	\ket{DL_N^{(m)}} = \left(^{N}_{m}\right)^{-1/2}\sum_k C_k \ket{0^{\otimes (N-m)}1^{\otimes m}}_k,
	\end{equation}
	and $p$ is the purity with which $\ket{DL_N^{(m)}}$ is created. 
	$\{k\}$ is the set of all distinct permutations of the spins. 
	In these states the $m$ excitations (occupations of state $\ket{1}$) are delocalized over all the qubits.
	Interestingly, these states are close to the maximally entangled Dicke states, that can be obtained when all the $C_k$ coefficients are equal.
	
	To create the state shown by Eqn.~\ref{eqn:state} we start with the state $\ket{0^{\otimes (N-m)}1^{\otimes m}}$, where the $m$ excitations are localized on the last $m$ qubits. Such a state can be easily achieved for arbitrary $m$ from a fully polarized state (all atoms in $\ket{0}$) if the two parts formed by first $N-m$ atoms and last $m$ atoms are separately addressable, which is also the minimal requirement to create the state in Eqn.~\ref{eqn:state}. Here one of the transverse beam shown in Fig.~\ref{fig:cartoon} addresses $m$ qubits and the other addresses $N-m$ qubits\footnote{This is the simplest case, in the tweezer setup all atoms can also be addressed individually.}. Once we have such a non-entangled state, the protocol shown in Fig.~\ref{fig:probability_evolution} then slowly delocalizes these $m$ excitations over all the qubits, creating an entangled state in the process. The system parameters we use are close to the ones required for two atoms undergoing quantum Zeno dynamics and adiabatic processes in a dressed state basis~\cite{wu_fast_2017}. Hence our protocol with off-resonant pulses should follow similar processes. The excitations hop from one atom to another via virtual excitations of the photons in the cavity. These processes help in keeping the system in the basis spanned by states $\ket{0^{\otimes (N-m)}1^{\otimes m}}_k$ as the total number of excitations are not changed. However the system is not perfect and in addition to this delocalization, there are incoherent processes that result in probability $(1-p)$ to occupy states where the excitations are different from $m$. The random nature of these processes mixes the target state $\ket{DL_N^{(m)}}$ with these other unwanted sates, the unwanted states are collectively denoted by $\hat{\rho}_\text{loss}$. These unwanted process are proportional to $\Gamma$ and $\kappa$ and modeled into the simulations by Lindblad master equation as shown in Eqn.~\ref{eqn:master_eqn}.
	
	\section{Quantum-enhanced Metrology}
	\label{sec_sensing}
	\begin{figure}
		\centering
		\includegraphics[width=0.9\textwidth]{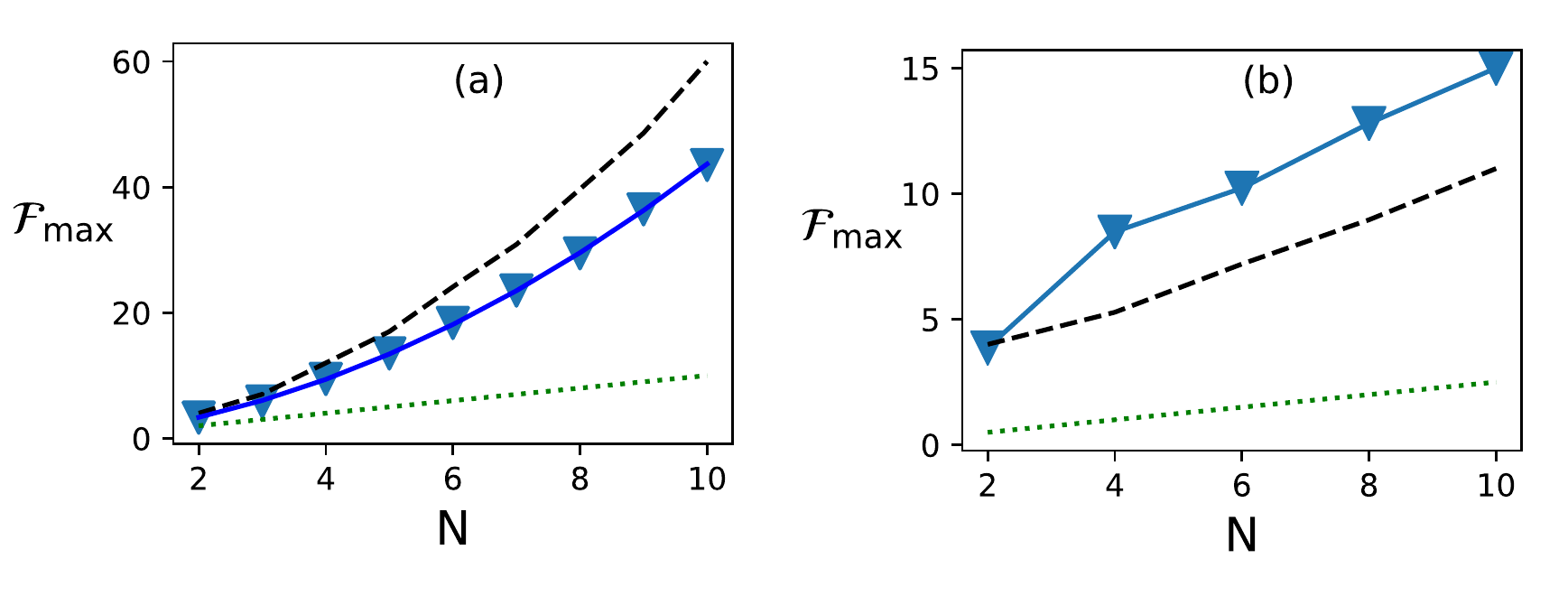}
		\caption{Maximum Fisher information ($\mathcal{F}_\mathrm{max}$) as a function of atom number for (a) case-I and (b) case-II. Dotted green line shows the quantum shot-noise limit, dashed black line is for Dicke states, blue triangles are for states we create. For (a) the blue solid line is a $N + a N^2$ fit to the blue triangles. For $N>6$, we use sequence optimized for $N = 6$.}
		\label{fig:sensitivity_vs_atno}
	\end{figure}
	In this section, we evaluate the metrological gain of the Dicke-like states~\cite{Toth_07} produced with our protocol for case I and II. 
	A change in magnetic field results in a change in the energy difference between the levels $\ket{0}$ and $\ket{1}$, which can be detected by measuring the change in the phase difference between these levels. Quantitatively, the effect of this field on the atoms can be represented by the following unitary matrix, 
	\begin{equation}
	\hat{U}_\text{ph} = e^{- i \hat{H}_\text{ph} },
	\end{equation}
	where $\hat{H}_\text{ph} = \sum_{A = 1}^{N} \theta_A \hat{\sigma}^A_z/2$ is the Hamiltonian generating the phase difference,  $\hat{\sigma}^A_z$ is the Pauli-$z$ operator for atom $A$, and  $\theta_A = f(B_A,t)$ is the angle that is the function of the magnetic field strength ($B_A$) for atom $A$ and the time of interaction ($t$). 
	
	For case I), after the state in Eqn.~\ref{eqn:state} is created, we first perform a $\pi/2$ rotation of each qubit individually,
	$\hat{U}_{\pi/2}\hat{\rho}_\text{DL}\hat{U}_{\pi/2}^{\dagger}$ and then we let it evolve under the action of $\hat{U}_\text{ph}$. 
	After a certain time $t$, $\theta(B,t)$ is imprinted on the state, and we then rotate it back with another $\pi/2$ pulse. After this stage a measurement of the atomic populations is performed. 
	As measurements can be performed on each atom separately in our system, each measurement will result in a projection onto the diagonal basis $\ket{b_j} = \ket{i_0,i_1...,i_N}$, where $i$ takes the values 0 and 1, and $N$ is the number of atoms. In Fig.~\ref{fig:sensitivity_vs_atno}(a) we report the Maximum Fisher information (maximized over $\theta$) as a function of the atom number for this case~\footnote{{Our computational resources allow us to perform the simulations only till N = 10. For $N>10$, the requirements for RAM and processor speeds are outside the available computational resources.}}. Here $\theta_A = \theta$ for all $A$.
	In Fig.~\ref{fig:sensitivity_vs_atno}(a), we also compare the Fisher information of our states with those of the fully symmetric Dicke states and the coherent states $\ket{1_\text{all}} =  \ket{1_1,1_2...,1_N}$. The non-entangled state gives a quantum shot-noise limit of $\mathcal{F}_\mathrm{max} = N$. We find that the states generated by our protocol perform better than the shot-noise limit and hence are useful for quantum metrology. A $N+a N^2$ fit to $\mathcal{F}_\mathrm{max}$ gives the $N^2$ component $a = 0.34\pm0.01$. The Dicke state's Fisher information scales instead as $\sim N + 0.5 N^2$~\cite{T_th_2014}. The reduction in $\mathcal{F}_\mathrm{max}$ for our states with respect to the ideal Dicke states is expected due to the presence of decoherence during the entanglement protocol. We also note that the optimal time evolution of fields found for two atoms is also optimal for $N>2$, removing the need to find an optimal evolution each time $N$ is changed. Assuming a best case scenario, extrapolating $\mathcal{F}_\mathrm{max}$ for our states with $N = 1000$ will lead to 25 dB gain in precision compared to classical systems. We note that the optimal time required to create such states scales linearly with $N$.    
	
	It is well known that the presence of correlated dephasing (common field fluctuations) greatly reduces the utility of a class of entangled states~\cite{Monz_14_2011} that are optimal for estimating a parameter which is common to all atoms. Hence, in presence of common noise, it is advantageous to create entangled states which are insensitive to such correlated dephasing~\cite{Dorner_2012}. The Dicke-like states ($\ket{DL_N^{(N/2)}}$) are such states, because of equal number of atoms in the state $\ket{0}$ and $\ket{1}$ for all of its components, all common phase changes are canceled out~\cite{Dorner_2012}. This makes them well suited to measure differences in fields or field gradients, which is the case-II). For case-II) the change in phase is such that $\theta_A = \theta/2$ for $A = 1$ to $N/2$ and $\theta_A = -\theta/2$ for $A = N/2+1$ to $N$, where $\theta$ is the phase difference introduced by a field difference $\Delta B$. 
	
	For case II) the Fisher information for a classical state is $\mathcal{F}_\mathrm{max} = N/4$, because each half of the atoms will detect the local field with $\mathcal{F}_\mathrm{max} = N/2$. Since we are interested in the difference of these local fields, $\mathcal{F}_\mathrm{max}$ decreases further by a factor of 2. From Fig.~\ref{fig:sensitivity_vs_atno}(b) we can see that the states we create (optimized for each $N$ separately) perform better than the quantum shot-noise limit also in this case. Notably, for case-II) our states outperform also the maximally entangled Dicke states in the region of parameters that we have explored.
	
\begin{figure}
		\centering
		\includegraphics[width=0.9\textwidth]{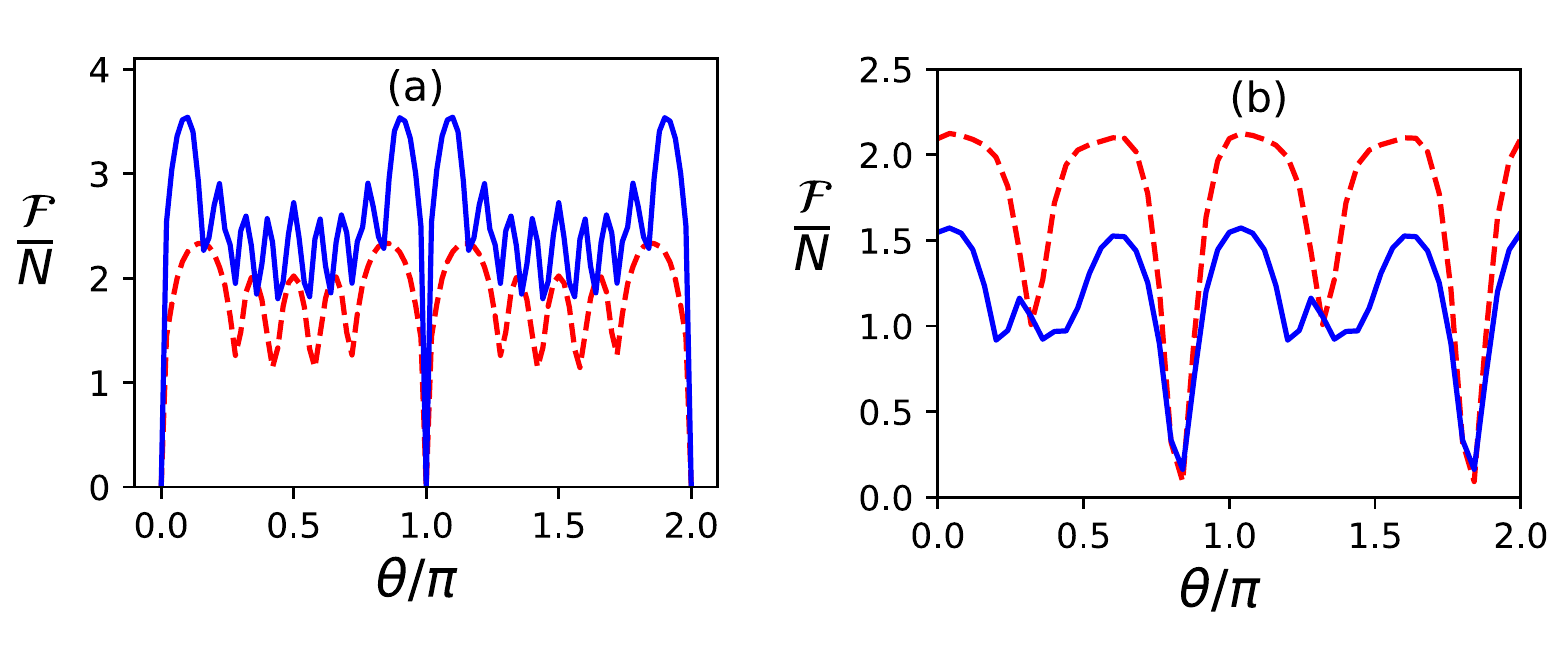}
		\caption{Fisher information as a function of the angle $\theta$. Dashed red line is for 4 atoms and solid blue line is for 8 atoms. (a) Case-I and (b) case-II.}
		\label{fig:sensitivity_vs_angle}
	\end{figure}

	The Fisher information for our Dicke-like states depends on the value of $\theta$. The scaling $\sim N + 0.5 N^2$ for the Dicke states is proven to hold for small $\theta$ and high $N$~\cite{T_th_2014}. We investigate the Fisher information of our scheme as a function of $\theta$, the results of which are shown in Fig.~\ref{fig:sensitivity_vs_angle}. We see that $\mathcal{F}$ oscillates as a function of $\theta$. These oscillations become less prominent as the number of atoms $N$ increases, as can be seen after comparing $\mathcal{F}$ for 4-atoms and 8-atoms from Fig.~\ref{fig:sensitivity_vs_angle}.   
	
	For our system, the maximum Fisher information is determined by the single atom cooperativity $\eta  = 4g_0^2/(\Gamma \kappa)$. Ideally one would want the photon-atom coupling $g_0$ to be considerably greater than the damping rates $\Gamma$ and $\kappa$ that introduce decohrence during the entangling operation. 
	One may decide to use atomic levels with small $\Gamma$; however this will also result in smaller $g_0$, which is not desirable. Also, the mode volume of the cavity can be reduced to increase the value of $g_0$ to as high as 215~MHz~\cite{colombe_strong_2007}. However, there are two disadvantages of this: the number of atoms which can fit in the cavity will reduce and the value of $\kappa$ will go up. Explicit calculations using the parameters form~\cite{colombe_strong_2007} showed no improvements in the Fisher information for our protocol.  
	
\begin{figure}
		\centering
		\includegraphics[width=0.5\textwidth]{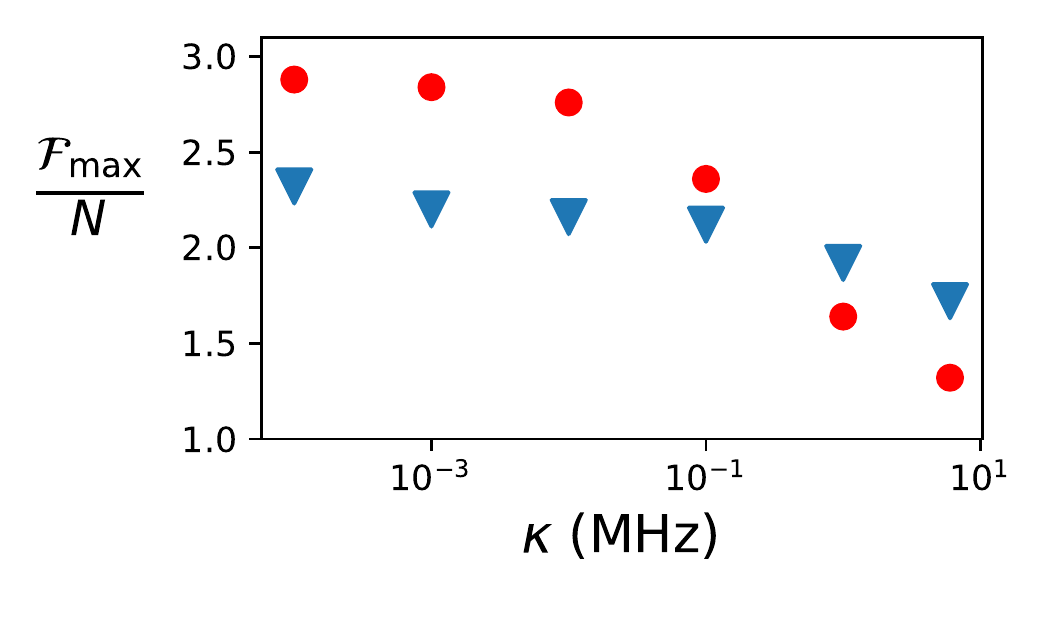}
		\caption{Fisher information as the function of $\kappa$. Red circles are for case-I and blue triangles for case-II. Here $\Delta = 0$, $\delta = 0$, $2\Gamma_0 = 2\Gamma_1 = 5.75$~MHz, $g_0 = 4$~MHz, and $N = 4$.}
		\label{fig:fid_vs_kappa}
	\end{figure}

	With better optical coatings for the cavity mirrors, we expect the value of $\kappa$ to reduce further and hence an improvement in the maximum Fisher information. In Fig.~\ref{fig:fid_vs_kappa} we compare the Fisher information for various values of $\kappa$ for case I) and case II), where we optimized the light pulses for each change in parameters. The results are shown in Fig.~\ref{fig:fid_vs_kappa}. As expected, the maximum Fisher information increases as $\kappa$ is reduced. The reduction flattens out after a $\kappa_\text{min}$. For case I) it occurs at $\kappa_\text{min} \approx 10^{-2}$ MHz and for case II) at $\kappa_\text{min} \approx 0.1$ MHz.
	
	\section{Conclusion}
	\label{conclusion}
	We have proposed a scheme for generation of multi-particle entanglement using an ensemble of singly-trapped atoms in an optical cavity. By using two external control fields, we can dynamically generate optimal entangled states according to the measurements one needs to perform. We have simulated a realistic system, accounting also for decoherence effects. We have performed these simulations for up to 10 atoms. Sources of entanglement infidelity were identified, and possible improvements in technology were suggested. We have shown how these entangled states can be used for performing quantum-enhanced metrology. We find that our proposed states perform better than the quantum shot-noise limit in measuring magnetic fields and magnetic field gradients. The protocol presented here is quite generic and can be used for other quantum sensing applications. For example, our states could be used to make better optical clocks. They also can be used to measure electric fields if one of the ground state is excited to a Rydberg state.   
    \ack
		The author was supported by the Leverhulme Trust Research Project Grant UltraQuTe (Grant No. RGP-2018-266) during this work. The author thanks Dr. Giovanni Barontini for feedback and guidance on this work. 
	\appendix	
	
		\section{Fisher information}
	\label{app:fish_info}
	The classical Fisher information is defined as,
	\begin{equation}
	\mathcal{F}(\theta)  =  \sum_{\mu} \frac{1}{P(\mu|\theta)} \left(\frac{\partial P(\mu|\theta)}{\partial \theta}\right)^2,
	\end{equation}
	where $P(\mu|\theta)$ is likelihood of getting an measurement result $\mu$ given the parameter $\theta$. In this article $P(\mu|\theta)  = p_j$, where $p_j$ is the diagonal element of density matrix at measurement stage such that we get a projection onto the basis vectors $\ket{b_j} = \ket{i_0,i_1...,i_N}$, where $i$ takes the values 0 and 1, $N$ is the number of atoms, and $j$ takes the values between 1 and $2^N$. $p_j(\theta) = \bra{b_j} \hat{U}_{\pi/2} \hat{U}_\text{ph}(\theta)\rho_0 \hat{U}_\text{ph}^{\dagger}(\theta)\hat{U}_{\pi/2}^{\dagger} \ket{b_j}$, where $\rho_0$ is the density matrix of the state used to measure the parameters $\theta$. Hence 
	\begin{equation}
	\begin{split}
	\frac{\partial p_j(\theta)}{\partial \theta} &= \frac{\partial \bra{b_j} \hat{U}_{\pi/2} \hat{U}_\text{ph}(\theta)\rho_0 \hat{U}_\text{ph}^{\dagger}(\theta)\hat{U}_{\pi/2}^{\dagger} \ket{b_j}}{\partial \theta} \\
	&= i \bra{b_j} \hat{U}_{\pi/2} [\rho_0,\hat{H}_\text{ph}(\theta)] \hat{U}_{\pi/2}^{\dagger} \ket{b_j},
	\end{split}
	\end{equation}
	where $[A,B]$ denotes the commutator of operators $A$ and $B$, and $\hat{H}_\text{ph}(\theta)$ is the Hamiltonian which generates the phase $\theta$. The Fisher information can be easily calculated if $\rho_0$ is known.

		\section{Hamiltonian and Hilbert space}\label{app:calculations}
		For $N$ atoms placed inside the cavity the interaction free Hamiltonian is given by,
		
		\begin{equation} 
		\begin{split}
		\hat{H}_0 =& \hbar\left[ \sum_{A = 1}^{N}\left(\omega_{1}\ket{1_A}\bra{1_A}+\omega_{e}\ket{e_A}\bra{e_A} \otimes \hat{I}_{\mathfrak{A}-A} \otimes \hat{I}_\text{c}\right) + \sum_{n}(\omega_\text{c}n\ket{n}\bra{n} \otimes \hat{I}_\mathfrak{A}) \right].
		\end{split}
		\end{equation}
		
		Here $\hat{I}_{\mathfrak{A}-A}$ is identity operator for all atoms minus the $A^\text{th}$ atom. $\hat{I}_\text{c}$ is identity operator for the cavity field, $n$ is cavity occupation number. $\hbar \omega_1 $ is energy of state $\ket{1}$, $\hbar \omega_e$ is energy of state $\ket{e}$ and $\omega_c$ is the resonant frequency of the cavity. This is combined Jaynes–Cumming~\cite{Jaynes_1963} Hamiltonian with rotating wave approximation for all atoms. 
		
		Atom-laser interaction Hamiltonian is,
		\begin{equation} 
		\hat{H}_\text{I}^\text{l} = \hbar \left[ \sum_{A = 1}^{N}\left((\Omega_A e^{i \omega_\text{l} t}\ket{0_A}\bra{e_A}+\Omega_A e^{-i \omega_\text{l}t}\ket{e_A}\bra{0_A}) \otimes \hat{I}_{\mathfrak{A}-A} \otimes \hat{I}_\text{c}\right) \right],
		\end{equation}
		
		where $\omega_\text{l}$ is the frequency of the laser addressing the atom, and $\Omega$ is its Rabi frequency. 
		The atom-cavity interaction Hamiltonian is,
		
		\begin{equation}
		\hat{H}_\text{I}^\text{c} = \hbar  \left[ \sum_{A = 1}^{N} (\sum_{n}\left(g_0 \sqrt{n+1}\ket{1_A,n+1}\bra{e_A,n}+ g_0\sqrt{n}\ket{e_A,n-1}\bra{1_A,n}) \otimes \hat{I}_{\mathfrak{A}-A}\right) \right].
		\end{equation}
		
		We convert the above Hamiltonians in the rotating frame of reference given by the following unitary operation for each atom,
		\begin{equation}
		\hat{U}_A =  \ket{0_A}\bra{0_A} +e^{i (\omega_\text{l}-\omega_\text{c})t} \ket{1_A}\bra{1_A}  +e^{i \omega_\text{l}t} \ket{e_A}\bra{e_A},
		\end{equation}
		and the following unitary for the cavity field,
		\begin{equation}
		\hat{U}_\text{c}=  \sum_{n} e^{i n \omega_\text{c} t } \ket{n}\bra{n}
		\end{equation}
		The combined unitary is,
		\begin{equation}
		\hat{U} =  \hat{U}_1 \otimes\hat{U}_2 .... \otimes \hat{U}_N \otimes \hat{U}_\text{c}
		\end{equation}
		
		In rotating frame we get, 
		\begin{equation} 
		\begin{split}
		\hat{H}_0 =& \hbar\left[ \sum_{A = 1}^{N}(\delta \ket{1_A}\bra{1_A}+\Delta \ket{e_A}\bra{e_A} \otimes \hat{I}_{\mathfrak{A}-A} \otimes \hat{I}_\text{c})  \right]. 
		\end{split}
		\end{equation}
		Here $\delta = \omega_1  - (\omega_\text{l} - \omega_\text{c})$ and $\Delta = \omega_e  - \omega_\text{l}$. 
		
		Atom-laser interaction,
		\begin{equation} 
		\hat{H}_\text{I}^\text{l} = \hbar \left[ \sum_{A = 1}^{N}((\Omega_A \ket{0_A}\bra{e_A}+\Omega_A \ket{e_A}\bra{0_A}) \otimes \hat{I}_{\mathfrak{A}-A} \otimes \hat{I}_\text{c}) \right].
		\end{equation}
		
		Atom-cavity interaction,
		
		\begin{equation}
		\hat{H}_\text{I}^\text{c} = \hbar  \left[ \sum_{A = 1}^{N} (\sum_{n}(g_0\sqrt{n+1}\ket{1_A,n+1}\bra{e_A,n}+ g_0\sqrt{n}\ket{e_A,n-1}\bra{1_A,n}) \otimes \hat{I}_{\mathfrak{A}-A}) \right].
		\end{equation}
		
		And the final Hamiltonian is then $\hat{H}_f = \hat{H}_0 + \hat{H}_\text{I}^\text{l}+\hat{H}_\text{I}^\text{c}$.
	
	For more than 6-atoms, the total Hilbert space of the system becomes prohibitively large for us to simulate. Hence we truncate the Hilbert space while solving Eqn.~\ref{eqn:master_eqn}. This truncated Hilbert space includes all the components of the state $\ket{DL_N^{(m)}}$, the cavity in state $\ket{0_c}$, all the other states which are connected to these states by the Hamiltonian $\hat{H}_f$ and the decay operators. These basis vectors are $P_k \ket{0^{\otimes (N-m)}1^{\otimes m}}\otimes \ket{0_c}$, the states $P_k \ket{0^{\otimes (N-m)}1^{\otimes (m-1)} e^1}\ket{0_c}$, $P_k \ket{0^{\otimes (N-m-1)}1^{\otimes m} e^1}\ket{0_c}$, $P_k \ket{0^{\otimes (N-m - 1)}1^{\otimes (m+1)}}\otimes \ket{1_c}$, and $P_k \ket{0^{\otimes (N-m - 1)}1^{\otimes (m+1)}}\otimes \ket{0_c}$. Here $P_k$ is the permutation operator. 

	After simulating Eqn.~\ref{eqn:master_eqn}, we replace the off-diagonal density matrix parts of basis vectors other than the components of the state $\ket{DL_N^{(m)}}$ to zero so that they do not contribute in the calculations of the Fisher information. 
	For 4-atoms we find that such a truncation gives the same result as the full Hilbert space, which validates the use of such truncation.

		\section{Atomic levels in magnetic field}\label{app:zeeman}
		\begin{figure}[h]
			\centering
			\includegraphics[width=0.64\textwidth]{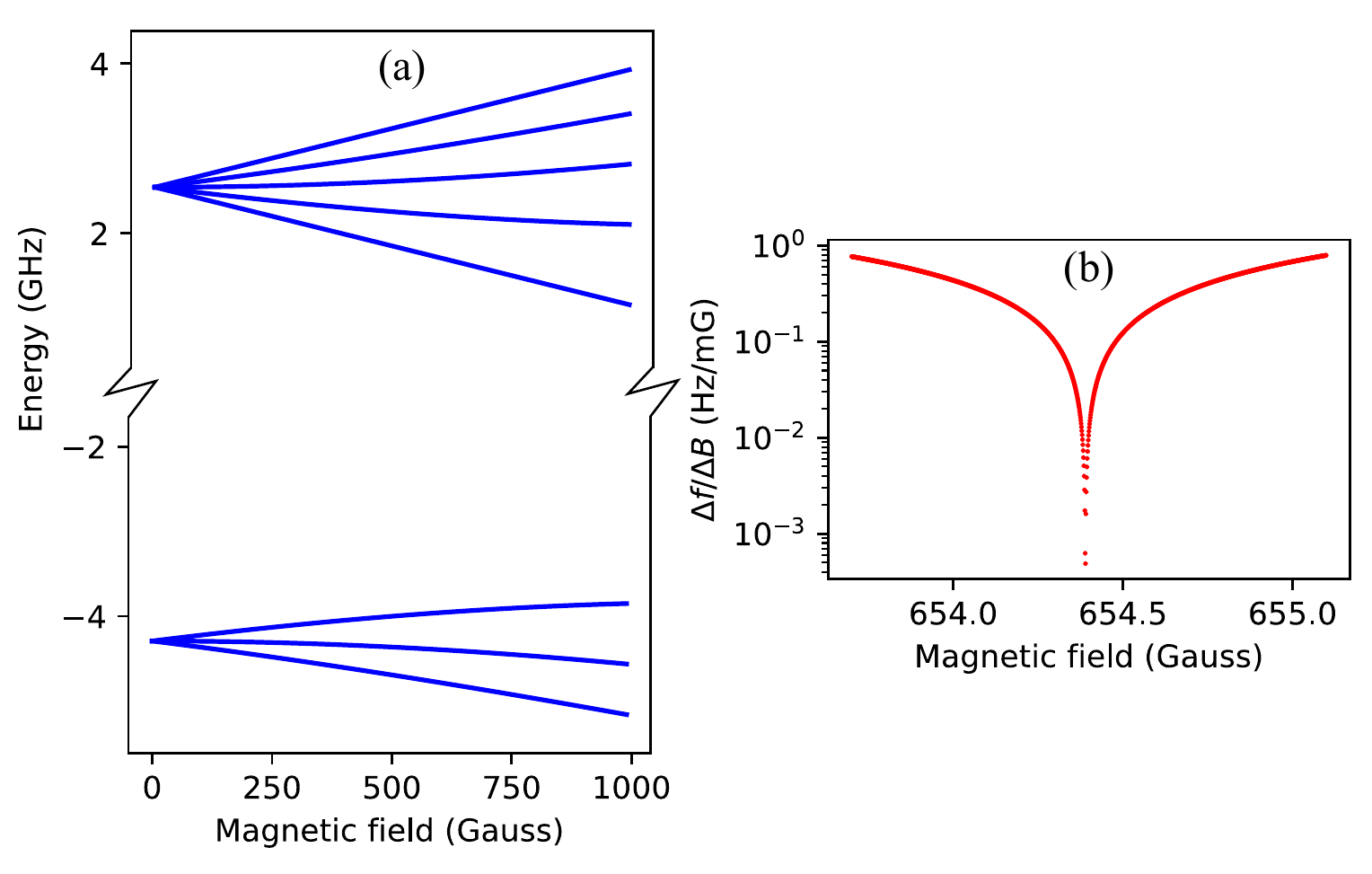}
			\caption{(a) Zeeman structure of the $5^{2}\text{S}_{1/2}$ state of the $^{87}$Rb atom. (b) Transition frequency fluctuation ($\Delta f$) for a 1~mG fluctuation of magnetic field ($\Delta B$) at different mean magnetic field strengths. This graph is for $(F,m_F) = (1,1)$ and $(F,m_F) = (2, 0)$ hyperfine levels of $5^{2}\text{S}_{1/2}$ state. Atomic constants for $^{87}$Rb are from ref.~\cite{Steck_Rb_87}.}
			\label{fig:levels_vs_Bfield}
		\end{figure}
		
\bibliography{biliography}
	
\end{document}